\documentclass[aip,jmp,amsmath,amssymb,preprint,]{revtex4-1}

\usepackage{graphicx}
\usepackage{dcolumn}
\usepackage{bm}
\usepackage[utf8]{inputenc}
\usepackage{url}
\usepackage{caption}

\usepackage{amsmath}
\newcommand{\RN}[1]{%
  \textup{\uppercase\expandafter{\romannumeral#1}}%
}

\draft
\begin{document}

\title{Suspended Spot-Size Converters for Scalable Single-Photon Devices}

\author{Asl{\i} D. U\u{g}urlu}
\email[]{Author to whom correspondence should be addressed: asli.ugurlu@nbi.ku.dk}
\affiliation{Center for Hybrid Quantum Networks (Hy-Q), Niels Bohr Institute, University of Copenhagen, Blegdamsvej 17, DK-2100 Copenhagen, Denmark}
\author{Henri Thyrrestrup}
\affiliation{Center for Hybrid Quantum Networks (Hy-Q), Niels Bohr Institute, University of Copenhagen, Blegdamsvej 17, DK-2100 Copenhagen, Denmark}
\author{Ravitej Uppu}
\affiliation{Center for Hybrid Quantum Networks (Hy-Q), Niels Bohr Institute, University of Copenhagen, Blegdamsvej 17, DK-2100 Copenhagen, Denmark}
\author{Claudéric Ouellet-Plamondon}
\affiliation{Center for Hybrid Quantum Networks (Hy-Q), Niels Bohr Institute, University of Copenhagen, Blegdamsvej 17, DK-2100 Copenhagen, Denmark}
\author{R{\"u}diger Schott}
\affiliation{Lehrstuhl f{\"u}r Angewandte Festk{\"o}rperphysik, Ruhr-Universit{\"a}t, Universitätsstrasse 150, D-44780 Bochum, Germany}
\author{Andreas D. Wieck}
\affiliation{Lehrstuhl f{\"u}r Angewandte Festk{\"o}rperphysik, Ruhr-Universit{\"a}t, Universitätsstrasse 150, D-44780 Bochum, Germany}
\author{Arne Ludwig}
\affiliation{Lehrstuhl f{\"u}r Angewandte Festk{\"o}rperphysik, Ruhr-Universit{\"a}t, Universitätsstrasse 150, D-44780 Bochum, Germany}
\author{Peter Lodahl}
\affiliation{Center for Hybrid Quantum Networks (Hy-Q), Niels Bohr Institute, University of Copenhagen, Blegdamsvej 17, DK-2100 Copenhagen, Denmark}
\author{Leonardo Midolo}
\affiliation{Center for Hybrid Quantum Networks (Hy-Q), Niels Bohr Institute, University of Copenhagen, Blegdamsvej 17, DK-2100 Copenhagen, Denmark}

\begin{abstract}
We report on the realization of a highly efficient optical spot-size converter for the end-face coupling of single photons from GaAs-based nanophotonic waveguides with embedded quantum dots. The converter is realized using an inverted taper and an epoxy polymer overlay providing a 1.3~$\mu$m output mode field diameter. We demonstrate the collection of single photons from a quantum dot into a lensed fiber with a rate of 5.84$\pm0.01$~MHz and estimate a chip-to-fiber coupling efficiency of $\sim48$~\%. The stability and compatibility with cryogenic temperatures make the epoxy waveguides a promising material to realize efficient and scalable interconnects between heterogeneous quantum photonic integrated circuits.
\end{abstract}

\pacs{Valid PACS appear here}   
\maketitle

\section{Introduction}
Single-photon devices such as emitters, routers, and detectors have been recently developed worldwide to extend the functionality of classical photonic integrated circuits towards on-chip quantum information processing [\citep{Dietrich2016}]. Chip-scale integration enables scaling up these devices to achieve multi-photon operation required for simulation and computing [\citep{Obrien2009,Rudolph2017}]. To this end, quantum dots (QDs) in gallium arsenide (GaAs) waveguides [\citep{Lodahl2015}] provide an excellent platform for generating photons with high indistinguishability and efficiency [\citep{Ding2016}]. However, despite the excellent progress in routing and detecting photons on this platform [\citep{Papon2019,Sprengers2011}], scaling quantum optics experiments with integrated circuits based exclusively on GaAs is rather challenging due to propagation loss. Interfacing QDs to a different material system could enable a whole new range of applications for ultra-fast switching, low-loss routing, and filtering. For example, in Refs. [\cite{Elshaari2017,Mnaymneh2019}] a nanowire QD single-photon source was integrated with silicon nitride waveguides enabling filtering and routing operation. Alternatively, GaAs waveguides have been epitaxially lifted-off, transferred to a SiN substrate and processed in-situ [\citep{Davanco2017,Bowers2016,Bovington2014}]. These approaches are challenging as they require compatible substrate materials (such as SiN) and introduce additional fabrication complexity when building electrical gates to QDs, which are essential to control charge noise of the emitter [\cite{Kuhlmann2013}].

An alternative approach exploits an optical device that allows for the end-fire coupling of single photons from the cleaved edge of a GaAs chip to another optical chip or into an optical fiber. Side-coupling from the chip edge requires suppressing back-reflections at the cleaved waveguide edge and, most importantly, enlarging the mode size to avoid field mismatch at the interface. While back-reflection can be controlled using anti-reflection coatings, angled cleaving, or by introducing inverted tapers [\cite{Cohen2013,Kirsanske2017}], the mode matching requires designing spot-size converters (SSC) to expand the optical mode profile. Inverted tapers with an overlay polymer are often adopted in silicon-on-insulator technology [\cite{Fan1999, McNab2003,Roelkens2006}] to perform this task. However, the fabrication process requires a low-index substrate (silica), which can physically support the polymer structure without introducing additional loss. Planar GaAs devices with deterministic photon-emitter coupling, on the contrary, are fabricated on a high-index sacrificial layer (Al$_x$Ga$_{1-x}$As) which is subsequently removed to form suspended waveguides [\cite{Midolo2015}]. An inverted taper for this platform has not yet been demonstrated.

We have developed a method to fabricate suspended converters to expand the optical spot size of a single-mode GaAs waveguide (refractive index $n=$ 3.48 and width 300~nm) to a Gaussian mode with 1.3~$\mu$m field diameter, which can be expanded further by increasing the converter length. The materials and processes used ensure vacuum and cryogenic compatibility and sufficient stability during cool-down and operation. The converter has an estimated efficiency of $\sim48$~\%, resulting in up to 5.84$\pm0.01$~MHz single-photon rates into a lensed optical fiber. In the present work a single lensed fiber is used  for characterization, but the developed SSC enables interfacing, e.g., multiple photonic chips as well, and can readily be scaled-up to multiple channels. As a further benefit of the fabricated device, it has been proposed that the polymer cladding may serve the dual purpose of efficiently suppressing limiting phonon decoherence processes [\cite{Dreesen2018}]. This paves the way for an on-demand fiber-coupled single-photon source with simultaneously near-unity degree of indistinguishability of the photons and near-unity coupling efficiency.

\begin{figure}
\begin{center}
\includegraphics{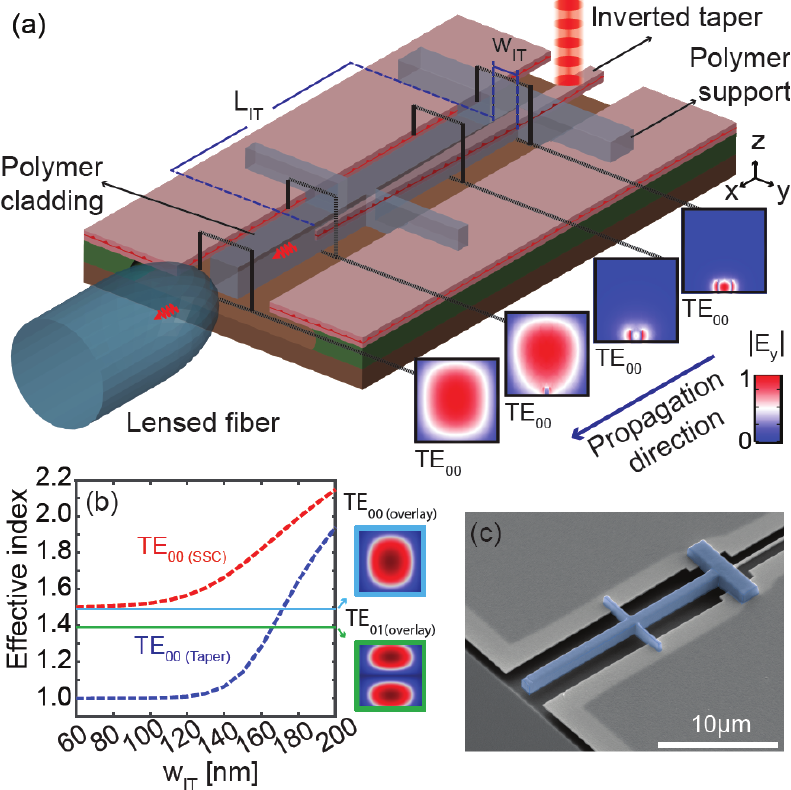}
\caption{\label{fig:concept} End-fire mode coupling between the SSC and a lensed fiber. (a) Sketch of the SSC made of a GaAs inverted taper and a polymer cladding. The inverted taper width and length are denoted by $w_{IT}$ and $L_{IT}$, respectively. A lensed fiber (with 1.3~$\mu$m mode-field diameter measured at $1/e^2$) is placed at the edge of the chip according to its working distance. The insets show the evolution of simulated mode profile cross-sections of the fundamental transverse electric (TE) mode as it propagates along the SSC. (b) Calculated effective refractive index of the fundamental TE modes of the SSC (red dotted line) and the GaAs inverted taper (blue dotted lines) as a function of $w_{IT}$. The light blue and green lines are effective indices of the fundamental and the first order TE modes of the overlay polymer waveguide, respectively (with their corresponding mode profiles). (c) Scanning electron microscope (SEM) image shows the fabricated spot-size converter. The highlighted blue shaded area is the overlay polymer.}
\end{center}
\end{figure}

The SSC presented here consists of a GaAs waveguide and an epoxy-based optical polymer cladding waveguide depicted schematically in Figure~\ref{fig:concept}(a). The width of the GaAs waveguide ($w_{IT}$) is gradually reduced to enlarge the area of its fundamental transverse electric (TE$_{00}$) mode [see inset of Figure~\ref{fig:concept}(a)], which is transferred to the overlay polymer waveguide with a size matched to the focal spot of a lensed fiber. The taper angle $\theta_t$ is designed to achieve an adiabatic transition into the fundamental mode of the cladding, according to $\theta_t \ll \Delta n_\mathrm{eff}$, where $ \Delta n_\mathrm{eff} $ is the difference between the effective indices between the first two modes of the cladding polymer. The adiabatic taper ensures a gradual decrease of the effective refractive index of TE$_{00}$ to match the low-index overlay epoxy polymer ($n_{Epo}$ = 1.52) and suppressing back-reflections in the taper. Figure~\ref{fig:concept}(b) shows the calculated effective index change in the fundamental modes of the SSC (red dotted line) and GaAs inverted taper (blue dotted line) as a function of the taper width ($w_{IT}$). The two horizontal lines indicate the effective indices of the fundamental and second-order (odd) mode, whose profiles are shown in the inset. The calculated difference $\Delta n_\mathrm{eff} \simeq 0.1$ requires a maximum taper angle well below 5~$^\circ$. It should be noted that larger mode diameters are in principle achievable with wider polymer waveguides, provided that smaller angles and longer tapers can be accurately patterned and fabricated. In the present work, we have chosen an angle below 1~$^\circ$ (taper length L$_{IT}$ around 10~$\mu$m), which is achievable with electron beam lithography and fulfills the adiabatic condition.

\subsection{Device fabrication}
\begin{figure}
\begin{center}
\includegraphics[width=8.5cm]{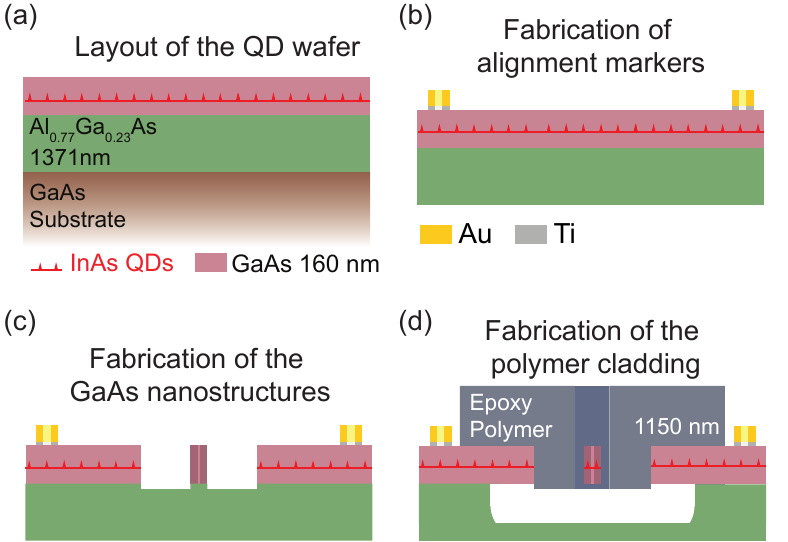}%
\caption{\label{fig:fab} Schematics of the fabrication process flow. (a) The layer structure of the wafer used for fabricating the SSC. (b) Ti/Au alignment markers are defined by e-beam evaporation and lift-off. (c) e-beam lithography followed by reactive ion etching (RIE) is used to define GaAs nanostructures on the membrane. (d) An epoxy polymer resin is spin coated on the substrate in order to for the overlay cladding and support structures by e-beam lithography.}%
\end{center}
\end{figure}

The suspended epoxy couplers, shown in the scanning electron microscope (SEM) image of Figure~\ref{fig:concept}(c), are fabricated on a GaAs wafer with the layer structure given in Figure~\ref{fig:fab}(a), grown by molecular beam epitaxy. The wafer contains a layer of self-assembled InAs QDs emitting at around 920~nm wavelength, located in the center of a 160-nm-thick membrane to ensure coupling to TE modes. Ti/Au marks are initially patterned and deposited on the wafer for precise alignment of the GaAs waveguides to the epoxy polymer (cf. Figure~\ref{fig:fab}(b)). The GaAs waveguide circuit is fabricated by means of electron beam (e-beam) lithography, followed by dry and wet etching [\cite{Midolo2015}]. The tapers are patterned on a 200-nm-thick CSAR 9 positive e-beam resist. A thin ($\sim 20$~${\buildrel _{\circ} \over {\mathrm{A}}}$) Ti layer is used as adhesion promoter to achieve narrow isolated lines of around 60 nm width. We have noticed that without proper adhesion, the resist tends to peel off at the tapered tip when the width is smaller than 100 nm. A slow-rate (35~nm/min) reactive ion etching (RIE) process (BCl3/Ar with 1:2 flow ratio) is used in combination with end-point detection to etch the membrane a few tens of nm into the sacrificial layer, cf. Figure~\ref{fig:fab}(c). This ensures that the spin-coating of the epoxy resin is sufficiently planar and that the bottom face of the epoxy waveguide is aligned to the GaAs waveguide.

A commercial epoxy polymer resin EpoCore2 (Microresist Gmbh) is used to coat the wafer with a final thickness of 1.15~$\mu$m. The polymer acts as a negative resist with sensitivity to both ultra-violet light (i-line) and electron beams [\citep{Del2007,Wahlbrink2009}]. Here, we used electron beam lithography in an Elionix F-125 system (acceleration voltage 125 keV and a uniform dose of 9~$\mu$C/cm$^2$) enabling smooth and vertical patterning of the waveguide with small ($<$ 20~nm) alignment errors. After patterning, the polymer is hard baked to make it adhere permanently on the substrate. Leftover residues from the epoxy are removed with an oxygen plasma step. Before undercut in a hydrofluoric acid solution (HF 5~\%), as shown in Figure~\ref{fig:fab}(d), the sample is cleaved to leave a flat polymer interface at the edge. Since the suspended structure can bend or collapse onto the substrate after the undercut due to capillary forces, the polymer is patterned with side tethers perpendicular to the propagation direction, in order to support it. A small bending, likely caused by strain relaxation, is observed on longer structures.

\subsection{Numerical analysis of the spot-size converter}
\begin{figure}
\begin{center}
\includegraphics{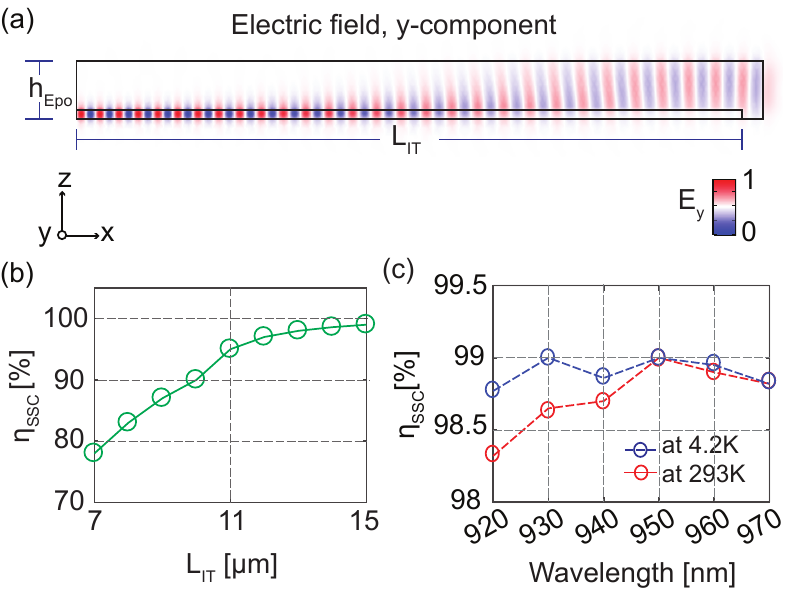}%
\caption{\label{fig:design} Numerical analysis of the SSC efficiency. (a) Cross-section of the 3D optical simulation showing the cut view of the inverted taper SSC, where the waveguide width is tapered from 200~nm to 60~nm over 15~$\mu$m length and the polymer height is 1.15~$\mu$m. (b) Calculated transmission efficiency $\eta_{SSC}$ as a function of the taper length. (c) Calculated transmission efficiency as a function of wavelength for material$'$ refractive indices at cryogenic (4.2 K) and room temperature (293 K) at $L_{IT}=15$~$\mu$m.}
\end{center}
\end{figure}

The design of the SSC is optimized by three-dimensional (3D) finite-element calculations. Figure~\ref{fig:design}(a) shows the distribution of the simulated $y$-component of the fundamental TE mode ($E_{y}$) plotted on the central cross-section of the device, and at a wavelength of 930~nm. The power transmitted over the SSC is evaluated using a scattering matrix (S-matrix) formalism to estimate the transmission efficiency of the SSC $\eta_{SSC}$. The efficiency depends on  the length of the linear inverted taper $L_{IT}$ and the width of the waveguide at the tip. As mentioned above, the smallest achievable tip in the fabrication process has a width of 60~nm, which is sufficient to achieve negligible mode mismatch (cf. the last two mode profiles in the inset of Figure~\ref{fig:concept}(a)). The transmission efficiency $\eta_{SSC}$ is plotted in Figure~\ref{fig:design}(b) as a function of the taper length. Near-unity ($\eta_{SSC}=98.9$~\%) is predicted when $L_{IT} = 15$~$\mu$m. Figure~\ref{fig:design}(c) shows the calculated efficiency for $L_{IT} = 15$~$\mu$m as a function of wavelength at room and cryogenic temperature. No significant deviation from the optimal coupling is observed over a 50~nm bandwidth, indicating that the coupler is very broadband. To avoid excessive bending of the structures after undercut, the samples are designed with 11~$\mu$m taper length, which results in slightly lower efficiency of $\eta_{SSC} = 94.6~\%.$

\begin{figure}
\begin{center}
\includegraphics[width=8.5cm]{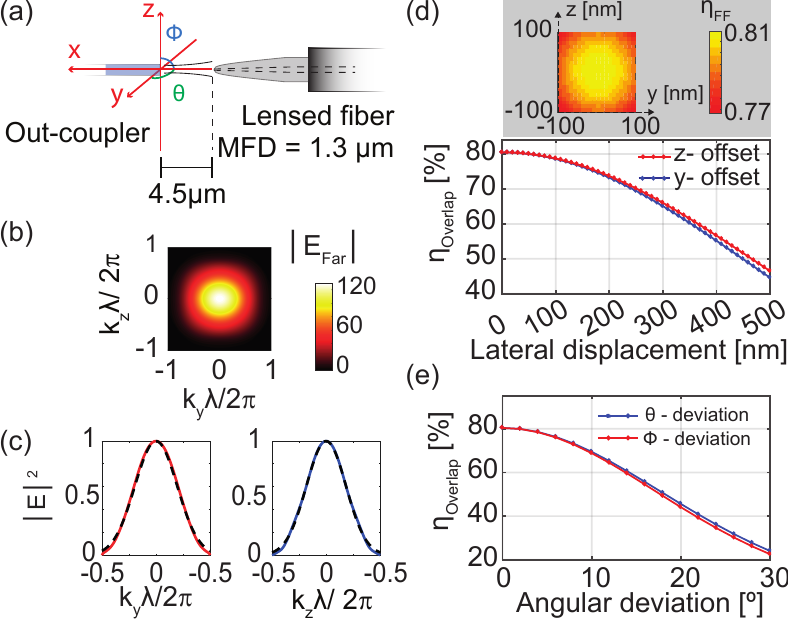}
\caption{\label{fig:farfield} Calculated mode overlap efficiency between the out-coupler and a lensed fiber. (a) Schematic of the optimal lateral alignment ($\Delta$x= 0, $\Delta$y=0) and angular adjustment($\Delta\theta$= 0, $\Delta\phi$=0) between the lensed fiber (MFD of 1.3~$\mu$m, measured at $1/e^2$ at a working distance of 4.5~$\mu$m) and the out-coupler. (b) Simulated far-field radiation pattern of the end-fire out-coupler. (c) Electric field distribution calculated along y-cut (red) and z-cut (blue) of (b), where the black dotted lines are the fitted Gaussian distributions of the lensed fiber in the focal plane. (d) Mode overlap efficiency $\eta_\mathrm{Overlap}$ as a function of lensed fiber lateral displacement in the y-direction (red) and z-direction (blue), when the out-coupler is positioned at the working distance of the fiber. The top inset shows the overlapping electric field map in the far field, where the facet of the out-coupler is scanned ($+$/$-$100~nm from the center in x and y) by the lensed fiber. (e) $\eta_\mathrm{Overlap}$ as a function of angular deviation ($\theta$ between x and y-axes, $\phi$ between x and z-axes) in the lensed fiber arrangement.}
\end{center}
\end{figure}

The SSC transmission efficiency $\eta_{SSC}$ is the theoretical maximum power transfer that the device would achieve in case of direct face-to-face coupling to a waveguide of same refractive index and mode profile (i.e., neglecting any additional index or mode mismatch related to the coupling). In this work, we have calculated the coupling efficiency for a lensed fiber with 1.3~$\mu$m mode-field diameter (MFD), which is positioned at a working distance of 4.5~$\mu$m from the end face of the chip. Such characterization method utilizing lensed fibers introduces additional loss related to the polymer-air-glass interfaces and to the alignment (displacement and angle) of the lensed fiber [the alignment scheme is illustrated in Figure~\ref{fig:farfield}(a)]. The plots in Figure~\ref{fig:farfield}(b) and \ref{fig:farfield}(c) show the calculated far-field distribution (two-dimensional map and cross-sections, respectively) of the optical mode emerging from the end face of the polymer, which appears nearly Gaussian.

We estimated the mode overlap by a normalized overlap integral between the complex electric field of the waveguide out-coupler ($E_{y,\mathrm{OC}}$) and the fiber ($E_{y,\mathrm{fiber}}$) as follows [\cite{Orobtchouka,Kataoka2010,AWSnyder2012}]:
\begin{equation}
	\eta_\mathrm{Overlap} = \frac{|\iint E_{y,\mathrm{oc}}^*E_{y,\mathrm{fiber}}\mathrm{d}y\mathrm{d}z|^2}{\iint |E_{y,\mathrm{oc}}|^2\mathrm{d}y\mathrm{d}z \cdot \iint |E_{y,\mathrm{fiber}}|^2\mathrm{d}y\mathrm{d}z},
	\label{eq:overlap_int}
\end{equation}
where the fundamental TE mode of the out-coupler ($E_{y,oc}$) is computed numerically and the circular Gaussian mode of the single mode lensed fiber ($E_{y,fiber}$) is approximated by a perfect Gaussian profile with an MFD of 1.3~$\mu$m. When the SSC is perfectly aligned to the focal plane of the lensed fiber the theoretical overlap efficiency is found to be 80.4~\%, indicating that the square profile of the epoxy waveguide introduces diffraction that is not captured by an ideal lensed fiber. Making wider polymer waveguides is expected to mitigate this issue.

The overlap integral of equation (\ref{eq:overlap_int}) allows us to estimate the alignment tolerance between the out-coupler and the lensed fiber. Figure~\ref{fig:farfield}(d) and Figure~\ref{fig:farfield}(e) show the coupling efficiency as a function of the lateral and angular fiber offset, respectively. We identify a 3 dB loss in the coupling efficiency for a lateral misalignment exceeding 0.5~$\mu$m or angular deviations around 25~$^\circ$. In our experiments, the fiber is mounted on nano-positioners with a sub-nm positioning resolution and 5~nm repeatability, which suggests that the lateral misalignment is not an issue. Angular deviations, however, are much more difficult to control as they require stages with rotational degrees of freedom. Additionally, the taper itself can slightly bend during cool-down due to thermal stress. We estimate that our total angle misalignment does not exceed 10~$^\circ$ by inspecting the setup with an optical microscope.

An additional loss term is caused by the reflection at the facet of the epoxy polymer out-coupler. We estimate them using the Fresnel coefficient for a polymer-air interface at normal incidence, resulting in approximately 4~\% loss ($R_\mathrm{polymer}=4$~\%). In summary, the theoretical chip-to-fiber coupling efficiency of the SSC has been estimated numerically to be 73~\% and is given by the product of the taper transmission efficiency $\eta_\mathrm{SSC} \simeq 94.6$~\%, the transmission across polymer-air interface $T_\mathrm{polymer}$= 1-$R_\mathrm{polymer}$ (96~\%), and the mode matching to a fiber, which in the ideal case is $\eta_\mathrm{Overlap}=80.4$~\%.


\section{Characterization and results}
\begin{figure}
\begin{center}
\includegraphics[width=8.5cm]{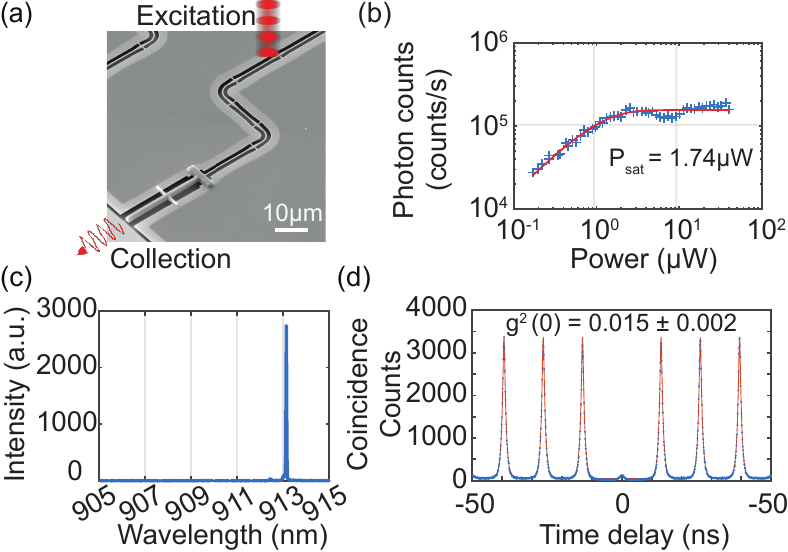}%
\caption{\label{fig:QD} Measurement of single-photon collection from the SSC. (a) Scanning electron microscope image of the nanobeam waveguide terminated by the SSC. The QD coupled to the waveguide is excited from the top by the pump laser and single photons are collected horizontally from the SSC at the edge of the sample. (b) Detected single-photon rate as a function of the pump power exciting the QD from which the saturation power is extracted. (c) Photoluminescence spectrum of a QD emitting at 913.1~nm under above-band excitation with a pulsed laser.  (d) Auto-correlation measurement of the QD emission line in (c) taken at a pump power of 1~$\mu$W. The coincidence count histogram is fitted with exponential decays in order to extract $g^{(2)}(0)=$~0.015$\pm$0.002.}%
\end{center}
\end{figure}

The sample is mounted on a piezo stack and cooled down to 4.2 K in a bath cryostat. A lensed fiber is mounted next to the sample on a separate piezo stack, fed through the cryostat using a hermetic sealing, and finally spliced to another single-mode fiber and interfaced to the detection setup. A microscope objective is used to excite the QDs from the top, and the photoluminescence (PL) is collected on the fiber by manually optimizing its alignment. We investigated a QD placed in a nanobeam waveguide, as shown in Figure~\ref{fig:QD}(a), terminated by an out-coupler. The QD line is pumped via above-band excitation using an 810~nm pulsed laser source with a repetition rate of 76~MHz. The QD integrated intensity as a function of the excitation power (from 0.2~$\mu$W to 40~$\mu$W) is shown in Figure~\ref{fig:QD}(b). The saturation power $P_\mathrm{sat} = 1.78\pm0.25~\mu$W is extracted at a 95~\% confidence interval by fitting the peak counts of the QD with the function $I = I_\mathrm{max}(1-\exp(-P/P_\mathrm{sat}))$ (solid red line). The spectrum of the QD is successively recorded by a cooled CCD camera (spectral resolution $\sim$ 20 pm). Figure~\ref{fig:QD}(c) shows the excitonic line of the QD for an excitation power P = 1~$\mu$W. Time-resolved PL (not shown) reveals two-time exponential decay, typical of neutral exciton emission. The fast decay rate, associated with the bright exciton, is 1.75~ns$^{-1}$ while the dark exciton recombination rate is 0.27~ns$^{-1}$, which is comparable to values reported previously in the literature [\citep{Johansen2010}].

In order to characterize the single-photon nature of the light emitted by the QD, we carried out an auto-correlation measurement using a Hanbury Brown-Twiss setup. The QD is pumped at 0.56x~$P_\mathrm{sat}$ and subsequently filtered with a grating setup (300 pm spectral bandwidth). The filtered signal is split by a 50:50 fiber beam-splitter whose outputs are connected to two fiber-coupled silicon avalanche photo-diodes (APDs). A time-tagging module (PicoHarp 300) is used to record correlation events between the two detectors. The result of the auto-correlation experiment is shown in Figure~\ref{fig:QD}(d), where the coincident counts are plotted as a function of the delay time between the two APDs. The plot reveals a strong anti-bunching at zero time delay, $g^{(2)}(0)= 0.015 \pm 0.002$, confirming the single-photon nature of the QD emission. The data are fitted with two-sided exponentials and the $g^{(2)}$ value is obtained by dividing the area of the central peak to the peak around a time delay of 250 ns. The purity of the source could be improved even further by fabricating the structures on a sample with lower QD density or by adopting different excitation schemes such as p-shell or resonant excitation [\cite{Somaschi2016}].

\section{Characterization of the coupling efficiency of the device}
The recorded single-photon emission rate from the QD can be further scrutinized in order to experimentally extract the efficiency of the SSC. We record in total a single-photon count rate of 167 kHz on the APD at the saturation.
Based on this number we can extract the photon rate inside the fiber to be $5.84\pm 0.01$~MHz by accounting for the measured transmission efficiency of the various components of the detection setup:  fiber splicing and mating sleeves: $\mathrm{T}$ = $32\pm2$~\%, grating setup efficiency: $\eta_\mathrm{F}$= $30\pm1$~\% and, detector quantum efficiency $\eta_\mathrm{APD}$= 30~\%. Consequently, the overall setup efficiency is $2.9\pm 0.3$~\%. Since the QD is triggered at a repetition rate of 76~MHz the overall source efficiency is found to be $7.7\pm 0.08$~\%, which is defined as the probability that a photon is collected by the fiber given that a QD is triggered. All on-chip and off-chip efficiencies are summarized in the Table~(\ref{tab:efficiencies}).

To get to the chip-to-fiber efficiency of the SSC we estimate the rate of single photons present inside the GaAs waveguide and compare that to the measured value of single photons in the fiber. 
Furthermore, since above-band excitation is applied, also additional exciton levels are occupied during excitation notably the non-radiative dark excitons leading to an effective preparation efficiency of the bright radiative exciton of $\eta_\mathrm{QD} \sim 50~\%$ [\citep{Johansen2010}]. Furthermore, since the waveguide is two-sided only half of the coupled photons are collected leading to  $\beta_\mathrm{directional}$ $\sim 40$~\% based on the calculated $\beta$-factor [\cite{Dreesen2018}].
Finally propagation loss in the 117~$\mu$m-long nanobeam waveguide amounts to $\eta_\mathrm{nb}$= $81\pm 2$~\% [\citep{Papon2019}], leading to an overall photon rate at the entrance of the SSC of 5.84$\pm0.01$~MHz.

Based on the above analysis, we can quantify the overall efficiency of the fiber-coupling comprising of both the SSC and the aligned tapered fiber. We estimate the chip-to-fiber coupling efficiency to be $\sim 48$~\%. The minor discrepancy between experimental estimate and theory can likely be attributed to the angular deviations in the alignment of the fiber or fabrication imperfections influencing the taper tip.

\begin{table}[h!]
\centering
\begin{tabular}{@{}ll@{}}
\RN{1}                                                &                         \\ 
\hline
\hline
Theoretical transmission efficiency $\eta_{SSC}$              & 94.6~\%                   \\
Theoretical polymer transmittance $T_\mathrm{polymer}$        & 96~\%                     \\
Theoretical mode overlap efficiency $\eta_\mathrm{Overlap}$   & 80.4~\%  					 \\
\hline
Theoretical coupling efficiency							      & 73~\%  					 \\
\hline
\hline
\RN{2}										         &                        \\
\hline
\hline
APD count rate                                       & 0.167~MHz               \\
\hline
Detector efficiency $\eta_\mathrm{APD}$              & 30~\% 		      \\
Spectral filter efficiency $\eta_\mathrm{F}$         & $30\pm1$~\%            \\
Collection optics efficiency $\mathrm{T}$            & $32\pm2$~\%            \\
\hline
Single-photon rate at the lensed fiber               & $5.84 \pm 0.01$~MHz    \\ 
\hline
\hline
\RN{3}												 &                       \\
\hline
\hline
Laser repetition rate                                & 76~MHz                \\
\hline
Preparation efficiency $\eta_\mathrm{QD}$            & $\sim 50$~\%           \\
Waveguide $\beta_\mathrm{directional}$				 & $\sim 40$~\% 	      \\	
Nanobeam transmission efficiency $\eta_\mathrm{nb}$  & $81\pm 2$~\%           \\
\hline
Single-photon rate at the SSC input                  & $\sim 12$~MHz      \\
\hline
\hline
\RN{4}												 &                        \\
\hline
\hline
Chip-to-fiber coupling efficiency                    & $\sim 48$~\% 	      \\		
\hline
\hline						
\end{tabular}
\caption{Transmission efficiencies and performances of the single-photon source out-coupled to a lensed fiber via the SSC. Part \RN{1}: theoretical coupling efficiency of the spot-size converter is a product of the SSC transmission efficiency (based on 3D FEM simulations), transmission efficiency of the polymer overlay waveguide as a result of reflection losses due to the Fresnel reflectivity at the cladding facet, mode overlap efficiency at far-field. Part \RN{2}: off-chip efficiencies for detecting, filtering, collection optics. Part \RN{3}: on-chip efficiencies of the QD excitation and single-photon transmission. Part \RN{4}: experimentally calculated chip-to-fiber coupling efficiency.}
\label{tab:efficiencies}
\end{table}

\section{Conclusions}
We reported the design, fabrication, and characterization of a suspended spot-size converter for the end-face coupling of single photons from QDs. The device holds great promise as a building block for scalable quantum information processing and for the realization of photonic quantum networks. The measured efficiency of $\sim 48$~\% is currently limited by the characterization method, which is based on a lensed fiber. Direct coupling to a second chip could lead to better alignment and would increase the single-photon rate further. Moreover, the design allows for the straightforward scaling of the number of couplers to multiple ports, thereby enabling multi-photon protocols. The efficiency of the device could readily be improved further: (i) termination of the waveguide on one side with a mirror will double the photon collection efficiency; (ii) resonant exciatation of the QD could boost the preparation efficiency to unity; (iii) longer waveguide tapers would improve the SSC efficiency. It is expected that any alignment imperfections of the fiber could be overcome chip-to-chip or chip-to-fiber coupling using the SSCs. For instance the SSC is readily applicable in a hybrid approach to quantum-information processing using two different photonic chips, where one is a source chip delivering deterministic photonic qubits and the other is a processing chip implementing, e.g., a complex quantum algorithm.


\begin{acknowledgments}
The authors gratefully acknowledge Xiaoyan Zhou for fruitful discussions about the numerical simulations and Zhe Liu for the helpful comments on the fabrication. We gratefully acknowledge financial support from the European Research Council (ERC Advanced Grant 'SCALE'), Danish National Research Foundation (Center of Excellence 'Hy-Q'), Innovation Fund Denmark (Quantum Innovation Center 'Qubiz'), Styrelsen for Forskning og Innovation (FI)(5072-00016B QUANTECH), Bundesministerium für Bildung und Forschung (BMBF) (16KIS0867, Q.Link.X), Deutsche Forschungsgemeinschaft (DFG) (TRR 160).
\end{acknowledgments}


\end{document}